\documentclass[pre,twocolumn,twoside,showpacs,byrevtex,superscriptaddress]{revtex4}

\usepackage{array}

\newcommand{\PreserveBackslash}[1]{\let\temp=\\#1\let\\=\temp}
\let\PBS = \PreserveBackslash

\usepackage{graphicx,epsfig,verbatim,enumerate}
\usepackage{amssymb,amsmath}
\usepackage{ifthen}
\newboolean{twocolswitch}

\usepackage{color}

\newcommand{\avg}[1]{\left\langle#1\right\rangle}
\newcommand{\tavg}[1]{\langle#1\rangle}

\newcommand{\sindex}[1]{}
\newcommand{\nindex}[1]{}

\newcommand{\etal}{\textit{et al.}}
\newcommand{\www}[1]{\url{#1}}

\newcommand{\req}[1]{(\ref{#1})}
\newcommand{\Req}[1]{Eq.~(\ref{#1})}

\newcommand{\veck}{\vec{k}}

\newcommand{\bidmark}{\rm u}
\newcommand{\inmark}{\rm i}
\newcommand{\outmark}{\rm o}

\newcommand{\kin}{k_{\inmark}}
\newcommand{\kout}{k_{\outmark}}
\newcommand{\kbid}{k_{\bidmark}}

\newcommand{\Probin}{P^{(\rm \inmark)}}
\newcommand{\Probout}{P^{(\rm \outmark)}}
\newcommand{\Probbid}{P^{(\rm \bidmark)}}

\setboolean{twocolswitch}{true}

\begin{document}

\title{
  Direct, physically-motivated derivation of the contagion condition
for spreading processes on generalized random networks

}

\author{
\firstname{Peter Sheridan}
\surname{Dodds}
}

\email{peter.dodds@uvm.edu}

\affiliation{Department of Mathematics \& Statistics,
  The University of Vermont,
  Burlington, VT 05401.}

\affiliation{Complex Systems Center
  \& the Vermont Advanced Computing Center,
  The University of Vermont,
  Burlington, VT 05401.}

\author{
\firstname{Kameron Decker}
\surname{Harris}
}

\email{kameron.harris@uvm.edu}

\affiliation{Department of Mathematics \& Statistics,
  The University of Vermont,
  Burlington, VT 05401.}

\affiliation{Complex Systems Center
  \& the Vermont Advanced Computing Center,
  The University of Vermont,
  Burlington, VT 05401.}

\author{
\firstname{Joshua L.}
\surname{Payne}
}

\email{joshua.payne@dartmouth.edu}

\affiliation{Computational Genetics Laboratory, 
                           Dartmouth College, Hanover, NH 03755}

\markboth{Title}
{Author names}

\date{\today}

\begin{abstract}
  For a broad range of single-seed contagion processes acting on generalized random networks,
we derive a unifying analytic expression for the possibility
of global spreading events in a straightforward, physically intuitive fashion.
Our reasoning lays bare a direct mechanical understanding 
of an archetypal spreading phenomena
that is not evident in circuitous extant mathematical approaches.

\end{abstract}


\maketitle

\section{Introduction}
\label{gcgrn.sec:introduction}

Spreading is a universal phenomenon occurring 
in many disparate systems across all scales, 
as exemplified by diffusion and wave propagation,
nuclear chain reactions,
the dynamics of infectious biological diseases
and computer viruses,
and the social transmission of religious and political beliefs.
Many spreading processes take place on networks,
or leave a branching network of altered entities 
in their wake,
and over the last decade, studies of contagion 
on random networks in particular have provided fundamental
insights through analytic results for abstract models~\cite{newman2003e,boguna2005a}.
Furthermore, in acknowledging the governing roles of 
the degree distribution~\cite{barabasi1999a}
and correlations between nodes~\cite{newman2003e},
generalized random networks~\cite{newman2003a} have been profitably
employed in modeling real-world networks~\cite{shen-orr2002a}.
Thus, a clear, physical understanding of the dynamics of contagion
processes on generalized random networks provides
a crucial analytic cornerstone for 
the goal of understanding spreading on real-world networks.

Here, we obtain a unifying analytic expression for the 
possibility of a global spreading event---which
we define as the infection of a non-zero fraction of an infinite network---for 
a broad range of contagion processes
acting on generalized random networks and starting from a single infected seed.
We provide both a general framework and
results for a series of specific random network families,
allowing us to reinterpret, integrate, and 
illuminate previously obtained conditions.
Our explanation has obvious pedagogic benefits:
While results for these families are known,
previous treatments have centered around powerful 
but non-intuitive and indirect mathematical approaches, 
typically involving probability generating functions~\cite{molloy1995a,newman2003a,boguna2005a}.
We show that a global spreading (or cascade) condition
can in fact be transparently derived by considering local growth rates 
of infection only, such that physical contagion processes are manifest in our expressions.

Our derivation readily accommodates networks
with an arbitrary mixture of
directed and undirected weighted edges,
node and edge characteristics,
and node-node correlations,
and can be extended to other kinds of random networks
such as bipartite affiliation graphs~\cite{newman2001b}.
Our argument also applies to contagion processes evolving
in continuous or discrete time, and for the latter case,
with either synchronous or asynchronous updates.
Nodes may also recover or stay infected 
as the outbreak spreads.

In what follows, we first obtain 
an inherently physical condition
for the possibility of spreading on generalized random networks,
and then provide specific treatments for
six interrelated classes of random networks.

\section{Physically-motivated derivation of a general spreading condition}
\label{gcgrn.sec:analysis}

Our goal is to intuitively derive a test for the possibility of global
spreading from a single seed, 
given a specific random network and contagion process~\cite{footnote:gcgrn.phase}.
To do so, we construct a global spreading condition based
on the infection counts of edge-node pairs rather than just nodes.
While considering how the number of infected nodes
grows is a more obvious and natural framing,
and one that has been broadly employed
(e.g., the reproduction number in mathematical epidemiology~\cite{murray2002a}),
the growth of `infected edges' emanating from infected nodes is equally transparent,
and opens a door to analytic treatment.

Since generalized 
random networks, correlated or not, are locally branching networks~\cite{newman2003a},
successful spreading from a single seed must entail nodes
becoming infected in response to a single neighbor's infection
(such nodes have been termed `vulnerable'~\cite{watts2002a}).
For any given contagion process, 
we therefore need only examine the transmission of infection along single edges.
Furthermore, successful spreading leads to exponential growth
on random networks when one infected edge, on average, generates
more than one new infected edge.

\begin{figure}[tp!]
  \centering
  \includegraphics[width=\columnwidth]{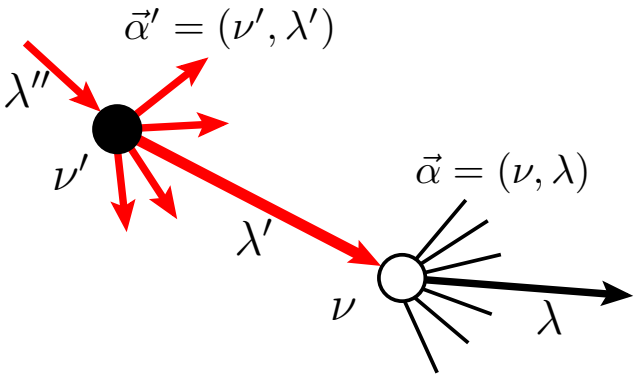}
  \caption{
    (Color online)
    Schematic showing an infection potentially spreading
    from node-edge pair 
    $\vec{\alpha}'=(\nu',\lambda')$ 
    to node-edge pair
    $\vec{\alpha}=(\nu,\lambda)$.
  }
  \label{fig:gcgrn.generalcontagion-schematic}
\end{figure}

In Fig.~\ref{fig:gcgrn.generalcontagion-schematic},
we provide a schematic of the spread
of a contagious element through a random network.
We frame our analysis around the 
probability that
an edge of type $\lambda'$ `infects an edge'
of type $\lambda$ through a node of type $\nu$,
where by type, we mean individual characteristics
such as node or edge age, node degree, edge direction, edge weight, hidden variables, etc.
As shown in Fig.~\ref{fig:gcgrn.generalcontagion-schematic}, a $\nu'$ node is already infected
due to a $\lambda''$ edge and is consequently
signalling its infection to its neighbors.
In particular, the $\lambda'$ edge communicates the infection
of the $\nu'$ node to the $\nu$ node 
and thereby potentially to the marked $\lambda$ edge.
For an infection to spread, we must account for
all possible edge-edge transitions incorporating
the probability of their occurrence based on 
(1) network structure and (2) the nature of the spreading process.
Our framing leads us to identify node-edge pairs 
as the key analytic components,
as indicated in Fig.~\ref{fig:gcgrn.generalcontagion-schematic},
and we write 
$\vec{\alpha}=(\nu,\lambda)$
and
$\vec{\alpha}'=(\nu',\lambda')$.

We first consider contagion processes with
discrete time updates and one-shot infection
chances.  By one-shot, we mean that once a node 
becomes infected, it has one time step to
infect its neighbors (excluding the node which infected it), 
after which no infection can be transmitted.
We argue that the growth of the 
expected number of 
type $\vec{\alpha}$ node-edge pairs
first infected at time $t$,
$f_{\vec{\alpha}}(t)$,
follows an exponential growth equation:
\begin{equation}
  f_{\vec{\alpha}}(t+1) 
  = 
  \sum_{\vec{\alpha}'}
  R_{\vec{\alpha} \vec{\alpha}'}
  f_{\vec{\alpha}'}(t),
  \label{eq:gcgrn.growthf}
\end{equation}
where 
$
R_{\vec{\alpha} \vec{\alpha}'}
$
is what we will call the `gain ratio matrix', 
and which possesses
a three-part form:
\begin{equation}
  \label{eq:gcgrn.Rgen}
  R_{\vec{\alpha} \vec{\alpha}'}
  =
  P_{\vec{\alpha} \vec{\alpha}'}
  \bullet
  k_{\vec{\alpha} \vec{\alpha}'}
  \bullet
  B_{\vec{\alpha} \vec{\alpha}'}.
\end{equation}
The first term 
$P_{\vec{\alpha} \vec{\alpha}'}$
represents the conditional probability that
a type $\lambda'$ edge emanating from a type $\nu'$ 
node leads to a type $\nu$ node.
The middle element
$k_{\vec{\alpha} \vec{\alpha}'}$
is the number of type $\lambda$ edges
emanating from nodes of type $\nu$,
excluding the incident type $\lambda'$ edge
arriving from a type $\nu'$ node.
The last term 
$B_{\vec{\alpha} \vec{\alpha}'}$
represents the probability that
a type $\nu$ node is infected by
a single infected type $\lambda'$ link arriving from a 
neighboring node of type $\nu'$
(the potential recovery of the infected $\nu'$ type node
is incorporated in $B_{\vec{\alpha} \vec{\alpha}'}$).
The first and second elements encode the network's structure,
while the third represents the spreading phenomenon,
and each term's dependence on $\vec{\alpha}$ and $\vec{\alpha}'$
may be none, part, or whole.
In \Req{eq:gcgrn.Rgen} and below,
we use the symbol `$\bullet$' to make clear the composition
of the three pieces of the gain ratio matrix.

We can now state the global spreading condition for spreading from
a single seed on arbitrarily correlated random networks with
discrete time update: the largest eigenvalue of
the gain ratio matrix $\mathbf{R} = [R_{\vec{\alpha} \vec{\alpha}'}]$ 
must exceed unity, i.e., 
\begin{equation}
  \sup
  \left\{
    |\mu| : \mu \in 
    \sigma
    \left(
      \textbf{R}
    \right)
  \right\}
  >
  1
  \label{eq:gcgrn.cascadecondition}
\end{equation}
where $\sigma(\cdot)$ indicates eigenvalue spectrum.

Next, we can easily accommodate other types of contagion processes
by computing the number of nodes infected a distance
$d$ away from the seed rather than as a function of time.  
The infection probability $B_{\vec{\alpha} \vec{\alpha}'}$
is then computed over all time and is interpreted
as the probability that a node of type $\nu$ is eventually
infected by edge $\lambda'$.  
We now more generally write
$
  f_{\vec{\alpha}}(d+1) 
  = 
  \sum_{\vec{\alpha}'}
  R_{\vec{\alpha} \vec{\alpha}'}
  f_{\vec{\alpha}'}(d),
$
with $B_{\vec{\alpha} \vec{\alpha}'}$'s role altered
and we see that the same 
global spreading condition arises.
Therefore, \Req{eq:gcgrn.cascadecondition}
applies for 
contagion processes for which time
is continuous or discrete, where nodes may recover,
etc., all providing we can sensibly compute
$B_{\vec{\alpha} \vec{\alpha}'}$~\cite{footnote:gcgrn.singleseed}.

\begin{table*}[tp!]
  \centering
  \begin{tabular}{|>{\PBS\raggedright\hspace{0pt}}m{0.13\textwidth}|>{\PBS\raggedright\hspace{2pt}}m{0.33\textwidth}|>{\hspace{3pt}}m{0.52\textwidth}|}
    \toprule
    \textbf{\, Network:} & 
    \textbf{\, Local Growth Equation:} &
    \textbf{\, Gain Ratio Matrix:} \\
    \colrule
    & & \\
    I. Undirected, Uncorrelated & 
    $f(d+1) = R f(d)$ &
    $
    \displaystyle
    R = 
    \sum_{\kbid}
    \Probbid(\kbid\, |\, \ast)
    \bullet
    (\kbid-1)
    \bullet
    B_{\kbid,\ast}
    $ \\
    & & \\
    II. Directed, Uncorrelated & 
    $f(d+1) = R f(d)$ &
    $
    \displaystyle
    R = 
    \sum_{\kin,\kout}
    \Probin(\kin,\kout\, |\, \ast)
    \bullet
    \kout
    \bullet
    B_{\kin,\ast}
    $ \\
    & & \\
    III. Mixed Directed and Undirected, Uncorrelated & 
    $
    \displaystyle
    \left[
    \begin{array}{c}
    f^{\rm (\bidmark)}(d+1) \\
    f^{\rm (\outmark)}(d+1)
    \end{array}
    \right]
    =
    \textbf{R} 
    \left[
    \begin{array}{c}
    f^{\rm (\bidmark)}(d) \\
    f^{\rm (\outmark)}(d)
    \end{array}
    \right]
    $ 
    &
    $
    \displaystyle
    \textbf{R} 
    = 
    \sum_{\veck}
    \left[
      \begin{array}{ll}
        \Probbid(\veck\,|\, \ast)
        \bullet
        (\kbid-1)
        &
        \Probin(\veck\,|\, \ast)
        \bullet
        \kbid
        \\
        \Probbid(\veck\,|\, \ast)
        \bullet
        \kout
        &
        \Probin(\veck\,|\, \ast)
        \bullet
        \kout
      \end{array}
    \right]
    \bullet
    B_{\kbid\kin,\ast}
    $ 
    \\
    & & \\
    \colrule
    & & \\
    IV. Undirected, Correlated & 
    $\displaystyle
    f_{\kbid}(d+1) = 
    \sum_{\kbid'}
    R_{\kbid\kbid'} 
    f_{\kbid'}(d)$ &
    $
    \displaystyle
    R_{\kbid\kbid'} =
    \Probbid(\kbid\, |\, \kbid')
    \bullet
    (\kbid-1)
    \bullet
    B_{\kbid\kbid'}
    $ \\
    & & \\
    V. Directed, Correlated & 
    $\displaystyle
    f_{\kin\kout}(d+1) = 
    \sum_{\kin',\kout'}
    R_{\kin\kout\kin'\kout'} 
    f_{\kin'\kout'}(d)$ 
    &
    $
    \displaystyle
    R_{\kin\kout\kin'\kout'} 
    =
    \Probin(\kin,\kout\, |\, \kin',\kout')
    \bullet
    \kout
    \bullet
    B_{\kin\kout\kin'\kout'} 
    $ \\
    & & \\
    VI. Mixed Directed and Undirected, Correlated & 
    $
    \displaystyle
    \left[
    \begin{array}{c}
    f_{\veck}^{\rm (\bidmark)}(d+1) \\
    f_{\veck}^{\rm (\outmark)}(d+1)
    \end{array}
    \right]
    =
    \sum_{k'}
    \textbf{R}_{\veck\veck'}
    \left[
    \begin{array}{c}
    f_{\veck'}^{\rm (\bidmark)}(d) \\
    f_{\veck'}^{\rm (\outmark)}(d)
    \end{array}
    \right]
    $ 
    &
    $
    \displaystyle
    \textbf{R}_{\veck\veck'}
    = 
    \left[
    \begin{array}{ll}
      \Probbid(\veck\,|\, \veck')
      \bullet
      (\kbid-1)
      &
      \Probin(\veck\,|\, \veck')
      \bullet
      \kbid
      \\
      \Probbid(\veck\,|\, \veck')
      \bullet
      \kout
      &
      \Probin(\veck\,|\, \veck')
      \bullet
      \kout
    \end{array}
    \right]
                \bullet
    B_{\veck\veck'}
    $ 
    \\
    & & \\
    \botrule
  \end{tabular}
  \caption{
    Summary of local growth equations and ratios
    for six classes of random networks and general contagion processes.
    These equations describe the expected early growth 
    in infection counts, represented by $f$, starting
    from a single initial infective (or seed).
    Each gain ratio is written so as to
    highlight three distinct factors
    in the following order:
    (1) the probability of an edge leading to a node of a specific type; 
    (2) the number of infected
    edges arising from a successful infection;
    (3) the probability of successful infection of that node.
    As the forms show and as discussed in the main text,
    these three factors depend on the nature of
    the edge potentially transmitting an infection.
    The gain ratio is a scalar for classes I and II,
    and a matrix for classes III--VI.
    When the gain ratio is a scalar, the contagion
    condition is simply $R > 1$, while for the matrix
    cases, at least one eigenvalue must exceed 1.
    Classes I--V are special cases of class VI.
    }
  \label{tab:gcgrn.cascadeconditions}
\end{table*}

\section{Application to undirected, directed, and mixed random networks}
\label{gcgrn.sec:mixed}

We now apply our argument to six interrelated
classes of random networks, connecting to existing
results in the literature.
We consider networks with arbitrary degree
distributions, mixtures of undirected and directed 
edges, and node-node correlations based on
node degree.  
Our general global spreading condition takes
on specific forms for these networks which
are worth deriving individually.
We summarize the resulting global
spreading conditions in Tab.~\ref{tab:gcgrn.cascadeconditions}.

We generally follow the approach of
Bogu\~{n}\'{a} and Serrano~\cite{boguna2005a},
who provided a formulation for 
degree-correlated random networks with
mixed undirected and directed edges.
We represent nodes by a degree vector
$\vec{k} = [\,\kbid\ \kin\ \kout\,]^{\rm T}$ where
the entries are, respectively,
the number of undirected (or bidirectional) edges between
a node and its neighboring nodes;
the number of directed edges leading in to a node;
and
the number of directed edges leading away from a node.
For random networks, the explicit inclusion of undirected edges
is necessary for modeling instances of mutual influence
between nodes, and analytically affords a way of connecting
directed networks with undirected ones.

We write the probability that a randomly selected node
has degree vector $\vec{k}$ as $P_{\vec{k}}$.
We represent correlations between
nodes via three transition probabilities: 
$\Probbid(\veck|\veck')$,
$\Probin(\veck|\veck')$,
and
$\Probout(\veck|\veck')$,
which are
the probabilities of an
undirected,
incoming, 
or 
outgoing
edge leading from 
a vector degree $\veck'$ node
to 
a vector degree $\veck$ node.
The superscripts therefore refer to the degree $\veck$ node
(these conditional probabilities are defined
similarly to those used in~\cite{boguna2005a}, but with
the directed cases reversed).

As we have argued in general,
in finding the global spreading condition for random networks, 
we have to determine three quantities:
(1) the probability that a type $\lambda'$ edge emanating from
an infected type $\nu'$ node leads to a 
type $\nu$ node where we may have to condition on 
$\vec{\alpha}'$ and $\vec{\alpha}$;
(2) in the case of successful infection,
the resultant number of newly infected outgoing 
$\lambda$ type edges emanating from 
the type $\nu$ node;
and
(3) the probability that the type $\nu$ node 
becomes infected.

We start with the basic case of undirected, uncorrelated random
networks with a prescribed degree distribution $P_{\kbid}$ (class I).
The first of the three quantities is 
given by the observation that following
a randomly chosen edge leads to a degree $\kbid$ 
node with probability $\kbid P_{\kbid} / \avg{\kbid}$~\cite{newman2001b},
which we will write as $\Probbid(\kbid\, |\, \ast)$
with the `$\ast$' indicating an absence of correlations.
Second, 
if a degree $\kbid$ node is infected,
$\kbid-1$ new edges will be infected.
And third, 
we have that a degree $\kbid$ node becomes
infected with probability $B_{\kbid,\ast}$.
Putting these pieces together and summing over
all possible values of $\kbid$ (since the network is uncorrelated),
we arrive at 
the well known global spreading condition for random networks:
\begin{equation}
  R 
  = 
  \sum_{\kbid=0}^{\infty} 
  \Probbid(\kbid\, |\, \ast)
  \bullet
  (\kbid-1)
  \bullet
  B_{\kbid,\ast}
  > 1.
  \label{eq:gcgrn.cc-prn}
\end{equation}
The local growth equation is simple: $f(d+1) = R f(d)$.
In the case that we set $B_{\kbid,\ast}=1$,
meaning the contagion process is always
successful, we have the condition for the 
presence of a giant component,
which was obtained
by Molloy and Reed~\cite{molloy1995a}
in the alternate form $\sum_{\kbid=0}^{\infty} \kbid (\kbid - 2) P_{\kbid} > 0$.
Although Molloy and Reed suggested some intuition for
this particular form, we believe the kind of derivation we
have provided here is the clearest, most direct formulation.
Later, Newman \etal~\cite{newman2001b} arrived at the
same result using generating functions,
specifically by examining when the average
size of finite components diverged for a family of
parametrized random networks,
and Watts~\cite{watts2002a}, using the same techniques,
obtained~\Req{eq:gcgrn.cc-prn} for a random network
version of Granovetter's threshold-based model of social contagion~\cite{granovetter1978a}.
These arguments, while entirely effective and 
part of a larger exploration of the details
of random networks (uncovering, for example, distributions of component sizes), 
are somewhat opaque and roundabout.
Thus, while we could readily rearrange~\Req{eq:gcgrn.cc-prn} and
our other results below to generate more mathematically clean
statements, an essential degree of physical intuition would be lost.

In moving to purely directed networks (class II), we now allow each node to have some
number of incoming and outgoing edges, $\kin$ and $\kout$.
The three pieces of the gain ratio $R$ are now:
(1) 
upon choosing a random (directed) edge, the probability the
edge leads to a node with degree vector $[\kin,\kout]^{\rm T}$ is 
 $\Probin(\kin,\kout\, |\, \ast) = k_i P_{\kin\kout} / \avg{\kin}$;
(2) 
the consequent number of infected outgoing edges is simply $\kout$;
and
(3)
the probability of infecting such a node is $B_{\kin,\ast}$.
The global spreading condition for uncorrelated directed networks is therefore
\begin{equation}
    R 
    = 
    \sum_{\kin,\kout}
    \Probin(\kin,\kout\, |\, \ast)
    \bullet
    \kout
    \bullet
    B_{\kin,\ast}
    > 1,
  \label{eq:gcgrn.cc-pdrn}
\end{equation}
and the local growth equation is again 
$f(d+1) = Rf(d)$.
The condition for the existence of a giant component, 
found by setting $B_{\kin,\ast}=1$ as before, 
was obtained by Newman \etal~\cite{newman2001b},
again by determining when the average size of
finite components diverges.
Newman \etal's version of the condition
is an elegant algebraic rearrangement of 
\Req{eq:gcgrn.cc-pdrn} as
$\tavg{2 \kout \kin - \kout -\kin} = 0$;
Bogu\~{n}\'{a} and Serrano~\cite{boguna2005a}
simplified \Req{eq:gcgrn.cc-pdrn} further to
$\tavg{\kout (\kin - 1)}=0$ since $\tavg{\kout}=\tavg{\kin}$.
Again, the physics of the process is 
entirely obscured by these 
mathematically clean statements.

We next consider random uncorrelated
networks with arbitrary mixtures of
directed and undirected edges (class III).
As shown in Tab.~\ref{tab:gcgrn.cascadeconditions},
the local growth equation now accounts for
the expected numbers of undirected and directed edges
a distance $d$ from the seed,
$f^{\rm (\bidmark)}(d)$
and 
$f^{\rm (\outmark)}(d)$
(outgoing rather than incoming edges are recorded
since we have framed our analysis around infected edges leaving
infected nodes).  
In computing the expected values of 
$f^{\rm (\bidmark)}(d+1)$
and 
$f^{\rm (\outmark)}(d+1)$,
we see the gain ratio is a 2$\times$2 matrix
built around four possible edge-edge transitions:
undirected to undirected,
undirected to outgoing,
incoming to undirected,
and 
incoming to outgoing.
The corresponding components
of the gain ratio matrix are 
$
\Probbid(\veck\,|\, \ast)
\bullet
(\kbid-1)
$,
$
\Probbid(\veck\,|\, \ast)
\bullet
\kout
$,
$
\Probin(\veck\,|\, \ast)
\bullet
\kbid
$,
and
$
\Probin(\veck\,|\, \ast)
\bullet
\kout.
$
For all four transitions, the probability
of infection is $B_{\kbid\kin,\ast}$.
Summing over all possible degrees,
we find global spreading occurs when the largest eigenvalue
of the gain rate matrix
\begin{equation}
  \textbf{R} 
  = 
  \sum_{\veck}
  \left[
    \begin{array}{ll}
      \Probbid(\veck\,|\, \ast)
      \bullet
      (\kbid-1)
      &
      \Probin(\veck\,|\, \ast)
      \bullet
      \kbid
      \\
      \Probbid(\veck\,|\, \ast)
      \bullet
      \kout
      &
      \Probin(\veck\,|\, \ast)
      \bullet
      \kout
    \end{array}
  \right]
          \bullet
  B_{\kbid\kin,\ast}
  \label{eq:gcgrn.mixedcond}  
\end{equation}
exceeds unity.  
The global spreading conditions
for pure undirected and directed networks,
Eqs.~\req{eq:gcgrn.cc-prn}
and 
\req{eq:gcgrn.cc-pdrn}
can be retrieved by setting
either 
$\Probbid(\veck\,|\, \ast)$
and
$\kbid$
or 
$\Probin(\veck\,|\, \ast)$ 
and 
$\kout$
equal to zero.

The above three classes of uncorrelated random networks
(I: undirected, II: directed, III: mixed) 
have natural degree-degree correlated versions (IV, V, VI).
The derivation of their respective global
spreading conditions follows the same argument
with two changes.  
First, averaging over node degrees
can no longer be done and the gain ratio matrix
now has entries for each possible transition
between edge types.  
Second, all transition probabilities are now
properly conditional, e.g.,
$\Probbid(\kbid\, |\, \ast)$ is replaced
with $\Probbid(\kbid\, |\, \kbid')$
for pure undirected random networks.
Consequently, the gain ratio matrix
is a function of the degrees $\veck'$ and $\veck$.
The resultant gain ratio matrices
and the expanded growth equations
agree with expressions obtained
by Bogu\~{n}\'{a} and Serrano~\cite{boguna2005a},
and are shown in Tab.~\ref{tab:gcgrn.cascadeconditions}

\section{Concluding remarks}
\label{gcgrn.sec:conclusion}

In summary, we have shown
that the possibility of global spreading 
for contagion processes on generalized random networks
can be obtained in a direct, physically motivated fashion.
A similar kind of clear approach should apply
for finding the probability of global spreading.
Our work naturally complements that of Gleeson and Cahalane~\cite{gleeson2007a}
who solved the fundamental problem of the final size
of an outbreak, in a similarly straightforward way
for macroscopic seeds and, in the limit, for isolated
seeds as well.
Obtaining an exact solution for the time evolution
of spreading from a single seed remains
the last major challenge for these random network models.

\begin{acknowledgments}
  PSD was supported by NSF CAREER Award \# 0846668;
JLP was supported by NIH grant \# K25-CA134286.

\end{acknowledgments}

\appendix


\begin{thebibliography}{13}
\expandafter\ifx\csname natexlab\endcsname\relax\def\natexlab#1{#1}\fi
\expandafter\ifx\csname bibnamefont\endcsname\relax
  \def\bibnamefont#1{#1}\fi
\expandafter\ifx\csname bibfnamefont\endcsname\relax
  \def\bibfnamefont#1{#1}\fi
\expandafter\ifx\csname citenamefont\endcsname\relax
  \def\citenamefont#1{#1}\fi
\expandafter\ifx\csname url\endcsname\relax
  \def\url#1{\texttt{#1}}\fi
\expandafter\ifx\csname urlprefix\endcsname\relax\def\urlprefix{URL }\fi
\providecommand{\bibinfo}[2]{#2}
\providecommand{\eprint}[2][]{\url{#2}}

\bibitem[{\citenamefont{Newman}(2003{\natexlab{a}})}]{newman2003e}
\bibinfo{author}{\bibfnamefont{M.~E.~J.} \bibnamefont{Newman}},
  \bibinfo{journal}{Phys. Rev. E} \textbf{\bibinfo{volume}{67}},
  \bibinfo{pages}{026126} (\bibinfo{year}{2003}{\natexlab{a}}).

\bibitem[{\citenamefont{Bogu\~{n}\'{a} and \'{A}ngeles
  Serrano}(2005)}]{boguna2005a}
\bibinfo{author}{\bibfnamefont{M.}~\bibnamefont{Bogu\~{n}\'{a}}}
  \bibnamefont{and} \bibinfo{author}{\bibfnamefont{M.}~\bibnamefont{\'{A}ngeles
  Serrano}}, \bibinfo{journal}{Phys. Rev. E} \textbf{\bibinfo{volume}{72}},
  \bibinfo{pages}{016106} (\bibinfo{year}{2005}).

\bibitem[{\citenamefont{Barab\'{a}si and Albert}(1999)}]{barabasi1999a}
\bibinfo{author}{\bibfnamefont{A.-L.} \bibnamefont{Barab\'{a}si}}
  \bibnamefont{and} \bibinfo{author}{\bibfnamefont{R.}~\bibnamefont{Albert}},
  \bibinfo{journal}{Science} \textbf{\bibinfo{volume}{286}},
  \bibinfo{pages}{509} (\bibinfo{year}{1999}).

\bibitem[{\citenamefont{Newman}(2003{\natexlab{b}})}]{newman2003a}
\bibinfo{author}{\bibfnamefont{M.~E.~J.} \bibnamefont{Newman}},
  \bibinfo{journal}{SIAM Review} \textbf{\bibinfo{volume}{45}},
  \bibinfo{pages}{167} (\bibinfo{year}{2003}{\natexlab{b}}).

\bibitem[{\citenamefont{Shen-Orr et~al.}(2002)\citenamefont{Shen-Orr, Milo,
  Mangan, and Alon}}]{shen-orr2002a}
\bibinfo{author}{\bibfnamefont{S.~S.} \bibnamefont{Shen-Orr}},
  \bibinfo{author}{\bibfnamefont{R.}~\bibnamefont{Milo}},
  \bibinfo{author}{\bibfnamefont{S.}~\bibnamefont{Mangan}}, \bibnamefont{and}
  \bibinfo{author}{\bibfnamefont{U.}~\bibnamefont{Alon}},
  \bibinfo{journal}{Nature Genetics} pp. \bibinfo{pages}{64--68}
  (\bibinfo{year}{2002}).

\bibitem[{\citenamefont{Molloy and Reed}(1995)}]{molloy1995a}
\bibinfo{author}{\bibfnamefont{M.}~\bibnamefont{Molloy}} \bibnamefont{and}
  \bibinfo{author}{\bibfnamefont{B.}~\bibnamefont{Reed}},
  \bibinfo{journal}{Random Structures and Algorithms}
  \textbf{\bibinfo{volume}{6}}, \bibinfo{pages}{161} (\bibinfo{year}{1995}).

\bibitem[{\citenamefont{Newman et~al.}(2001)\citenamefont{Newman, Strogatz, and
  Watts}}]{newman2001b}
\bibinfo{author}{\bibfnamefont{M.~E.~J.} \bibnamefont{Newman}},
  \bibinfo{author}{\bibfnamefont{S.~H.} \bibnamefont{Strogatz}},
  \bibnamefont{and} \bibinfo{author}{\bibfnamefont{D.~J.} \bibnamefont{Watts}},
  \bibinfo{journal}{Phys. Rev. E} \textbf{\bibinfo{volume}{64}},
  \bibinfo{pages}{026118} (\bibinfo{year}{2001}).

\bibitem[{foo({\natexlab{a}})}]{footnote:gcgrn.phase}
\bibinfo{note}{If we have a parametrized family of networks and contagion
  processes then we will be able to identify a phase transition between
  non-spreading and spreading.}

\bibitem[{\citenamefont{Murray}(2002)}]{murray2002a}
\bibinfo{author}{\bibfnamefont{J.~D.} \bibnamefont{Murray}},
  \emph{\bibinfo{title}{Mathematical Biology}} (\bibinfo{publisher}{Springer},
  \bibinfo{address}{New York}, \bibinfo{year}{2002}),
  \bibinfo{edition}{{T}hird} ed.

\bibitem[{\citenamefont{Watts}(2002)}]{watts2002a}
\bibinfo{author}{\bibfnamefont{D.~J.} \bibnamefont{Watts}},
  \bibinfo{journal}{Proc. Natl. Acad. Sci.} \textbf{\bibinfo{volume}{99}},
  \bibinfo{pages}{5766} (\bibinfo{year}{2002}).

\bibitem[{foo({\natexlab{b}})}]{footnote:gcgrn.singleseed}
\bibinfo{note}{Our approach does not explicitly require an initial single seed,
  and indeed randomly distributed isolated seeds possess the same global
  spreading condition, \Req{eq:gcgrn.cascadecondition}. If the seeds however
  constitute a non-zero fraction of the network, then the results
  of~\cite{gleeson2007a} apply.}

\bibitem[{\citenamefont{Granovetter}(1978)}]{granovetter1978a}
\bibinfo{author}{\bibfnamefont{M.}~\bibnamefont{Granovetter}},
  \bibinfo{journal}{Am. J. Sociol.} \textbf{\bibinfo{volume}{83}},
  \bibinfo{pages}{1420} (\bibinfo{year}{1978}).

\bibitem[{\citenamefont{Gleeson and Cahalane}(2007)}]{gleeson2007a}
\bibinfo{author}{\bibfnamefont{J.~P.} \bibnamefont{Gleeson}} \bibnamefont{and}
  \bibinfo{author}{\bibfnamefont{D.~J.} \bibnamefont{Cahalane}},
  \bibinfo{journal}{Phys. Rev. E} \textbf{\bibinfo{volume}{75}},
  \bibinfo{pages}{056103} (\bibinfo{year}{2007}).

\end{thebibliography}
\end{document}